\documentclass[10pt]{article}
\usepackage{graphicx}
\usepackage{amsmath}
\usepackage{amssymb}
\usepackage{caption2}
\begin{document}
\begin{center}
{\large {\bf \sc{Strong decays of the bottom mesons $B_1(5721)$,   $B_2(5747)$, $B_{s1}(5830)$, $B_{s2}(5840)$ and $B(5970)$ }}} \\[2mm]
Zhi-Gang Wang  \footnote{E-mail:zgwang@aliyun.com. } \\
  Department of Physics, North China Electric Power University, Baoding 071003, P. R.
  China
\end{center}

\begin{abstract}
In this article, we study the two-body strong decays of the bottom mesons with the heavy meson  effective theory in the leading order approximation,
and obtain all the analytical expressions of the decay widths of the light pseudoscalar meson transitions  among the S-wave, P-wave and D-wave bottom mesons.
  As an application, we tentatively assign  the  bottom meson $B(5970)$ as the $2{\rm S}\,1^-$, $1{\rm D}\,1^-$ and $1{\rm D}\,3^-$ states, respectively,
and calculate the decay widths of the $B_1(5721)$,   $B_2(5747)$, $B_{s1}(5830)$, $B_{s2}(5840)$ and $B(5970)$, which can be confronted with the experimental data in the future.
\end{abstract}


\section{Introduction}
The orbitally excited $B$-mesons $B_J^*(5732)$ or $B^{**}$ were   firstly observed by the  DELPHI, OPAL and  ALEPH collaborations in
electron-positron collisions at the Large Electron-Positron Collider  (LEP) \cite{LEP}.
In 2007, the D0 collaboration firstly observed the $B_1(5721)^0$ and $B_2(5747)^0$, and
measured the masses $M_{B_1}= (5720.6 \pm 2.4 \pm 1.4) \,\rm{MeV}$ and $M_{B_2^*}= (5746.8 \pm 2.4 \pm 1.7) \,\rm{MeV}$  \cite{D0-2007}.
Later, the CDF   collaboration confirmed  the $B_1(5721)^0$ and $B_2(5747)^0$, and
measured the masses $M_{B_1}= \left( 5725.3^{+1.6}_{-2.2} {}^{+1.4}_{-1.5} \right) \,\rm{MeV}$, $M_{B_2^*}= \left(5740.2^{+1.7}_{-1.8} {}^{+0.9}_{-0.8}\right) \,\rm{MeV}$, and obtained  the width $\Gamma_{B_2^*}= \left(22.7^{+3.8}_{-3.2} {}^{+3.2}_{-10.2}\right)\,\rm{MeV}$ for the first time \cite{CDF-2008}.
Also in 2007, the CDF  collaboration observed the $B_{s1}(5830)$ and $B_{s2}^*(5840)$, and measured the masses $M_{B_{s1}}=(5829.4 \pm 0.7)\,\rm{ MeV}$ and $M_{B_{s2}^*}=(5839.6 \pm 0.7)\,\rm{MeV}$ \cite{CDF-Bs-2007}.
The D0 collaboration confirmed the  $B_{s2}^*(5840)$ and obtained the mass $M_{B_{s2}^*}=(5839.6 \pm 1.1 \pm 0.7)\,\rm{MeV}$ \cite{D0-Bs-2007}.
In 2012, the LHCb collaboration updated the masses $M_{B_{s1}}=(5828.40\pm 0.04\pm 0.04\pm 0.41)\,\rm{ MeV}$ and $M_{B_{s2}^*}=(5839.99\pm 0.05\pm 0.11\pm 0.17)\,\rm{MeV}$ \cite{LHCb-2012}.

Recently, the CDF collaboration  reported the first evidence for a new resonance $B(5970)$ in the $B^+\pi^-$ and $B^0 \pi^+$ mass distributions with a significance
of $4.4\sigma$, and measured the masses $M_{B(5970)^0}=(5978 \pm 5\pm 12)\,\rm{ MeV}$, $M_{B(5970)^+}=(5961 \pm 5 \pm 12)\,\rm{ MeV}$,
and the widths   $\Gamma_{B(5970)^0}=(70 \pm 18 \pm 31)\,\rm{ MeV}$, $\Gamma_{B(5970)^+}=(60 \pm 20 \pm 40)\,\rm{ MeV}$ \cite{CDF-2013}.

There have been several approaches to calculate the masses of the established  bottom mesons $B_1(5721)$,   $B_2(5747)$, $B_{s1}(5830)$, $B_{s2}(5840)$, such as the
heavy quark effective theory \cite{HQEF-mass}, lattice QCD \cite{LQCD-mass}, potential models \cite{Potential-mass,EFG,PE}, heavy quark symmetry \cite{HQS-mass},
heavy meson chiral theory \cite{HMCh-mass}, QCD string model \cite{QCD-string-mass}, etc. The theoretical values  vary in the range $M_{\rm {exp}}\pm (50-100)\,\rm{MeV}$ \cite{HQEF-mass,LQCD-mass,Potential-mass,EFG,PE,HQS-mass,HMCh-mass,QCD-string-mass}.
Although the mass is a fundamental parameter in describing a hadron, the mass alone cannot validate the assignment.
For example, now the doublet $(D_{s0}^*(2317),D_{s1}(2460))$ is widely accepted to be the $\rm{1P}$ $(0^+,1^+)_{\frac{1}{2}}$ doublet, however, the masses of the $D_{s0}^*(2317)$ and $D_{s1}(2460)$ lie below the predictions of the potential models about $(100-150)\,\rm{MeV}$  \cite{Review-D}.

In Table 1, we present the predictions  from two typical potential models compared to the experimental data \cite{EFG,PE,PDG}.
The CDF collaboration  observed the  $B(5970)$ in the strong decays $B(5970) \to B^+\pi^-$, $B^0 \pi^+$ \cite{CDF-2013}, the possible quantum numbers are $J^P=0^+,\,1^-,\,2^+,\,3^-,\,4^+,\,5^-$, $\cdots$.
We can assign  the $B(5970)$ as the $2\rm{S}\,1^-$, $1\rm{D}\,1^-$ and $1\rm{D}\,3^-$ states tentatively according to the masses, see Table 1. In Ref.\cite{Zhu1401}, Y.Sun et al take the $B(5970)$ as the $2\rm{S}\,1^-$ state, and calculate the strong decays of the bottom mesons with the ${}^{3}P_0$ model.

In Refs.\cite{Wang1009,Wang1308}, we study the strong decays of  the charmed mesons $D(2550)$, $D(2600)$, $D(2750)$, $D(2760)$, $D_J(2580)$, $D_J^*(2650)$, $D_J(2740)$,
$D^*_J(2760)$, $D_J(3000)$, $D_J^*(3000)$ with
the heavy meson effective theory in the leading order approximation, and calculate  the decay widths and the ratios among the decay widths.
The ratios can be compared to the experimental data in the future
 to distinguish the different assignments. The heavy meson  effective theory have been applied
to identify the charmed  mesons \cite{Wang1009,Wang1308,Colangelo1207,Colangelo0511,Colangelo0607,Colangelo0710,Colangelo1001}, and
to calculate the radiative, vector-meson, two-pion decays
of the heavy quarkonium states \cite{HMET-RV}.

In this work, we study the two-body strong decays of the bottom mesons with the heavy meson  effective theory in the leading order approximation,
and obtain all the analytical expressions of the decay widths among the S-wave, P-wave and D-wave bottom mesons,
and calculate the decay widths of the $B_1(5721)$,   $B_2(5747)$, $B_{s1}(5830)$, $B_{s2}(5840)$ and $B(5970)$, which can be compared to the experimental data in the future.

\begin{table}
\begin{center}
\begin{tabular}{|c||c|c|c|c|c|c|c| }\hline\hline
$n\,\,L\,\, s_\ell\,\,    J^P$        &$b\bar{q}$ \cite{EFG} &$b\bar{q}$ \cite{PE} &\cite{PDG}        &$b\bar{s}$ \cite{EFG} &$b\bar{s}$ \cite{PE} &\cite{PDG} \\ \hline
$1\,\,{\rm S}\,\,\frac{1}{2}\,\,0^-$  &5280                  &5279                 &$5279.55\pm0.26$  &5372                  &5373                 &$5366.7\pm0.4$\\
$1\,\,{\rm S}\,\,\frac{1}{2}\,\,1^-$  &5326                  &5324                 &$5325.2\pm0.4$    &5414                  &5421                 &$5415.8\pm1.5$  \\

$2\,\,{\rm S}\,\,\frac{1}{2}\,\,0^-$   &5890                 &5886                 &--                 &5976                 &5985                 &--\\
$2\,\,{\rm S}\,\,\frac{1}{2}\,\,1^-$   &5906                 &5920                 &?5970              &5992                 &6019                 &--\\

$3\,\,{\rm S}\,\,\frac{1}{2}\,\,0^-$   &6379                 &6320                 &--                 &6467                 &6421                 &--\\
$3\,\,{\rm S}\,\,\frac{1}{2}\,\,1^-$   &6387                 &6347                 &--                 &6475                 &6449                 &--\\

$1\,\,{\rm P}\,\,\frac{1}{2}\,\,0^+$   &5749                 &5706                 &--                 &5833                 &5804                 &--\\
$1\,\,{\rm P}\,\,\frac{1}{2}\,\,1^+$   &5774                 &5742                 &--                 &5865                 &5842                 &--\\
$1\,\,{\rm P}\,\,\frac{3}{2}\,\,1^+$   &5723                 &5700                 &$5723.5\pm2.0$     &5831                 &5805                 &$5828.7\pm0.4$ \\
$1\,\,{\rm P}\,\,\frac{3}{2}\,\,2^+$   &5741                 &5714                 &$5743\pm5$         &5842                 &5820                 &$5839.96\pm0.20$  \\

$2\,\,{\rm P}\,\,\frac{1}{2}\,\,0^+$   &6221                 &6163                 &--                 &6318                 &6264                 &-- \\
$2\,\,{\rm P}\,\,\frac{1}{2}\,\,1^+$   &6281                 &6194                 &--                 &6345                 &6296                 &--\\
$2\,\,{\rm P}\,\,\frac{3}{2}\,\,1^+$   &6209                 &6175                 &--                 &6321                 &6278                 &-- \\
$2\,\,{\rm P}\,\,\frac{3}{2}\,\,2^+$   &6260                 &6188                 &--                 &6359                 &6292                 &-- \\

$1\,\,{\rm D}\,\,\frac{3}{2}\,\,1^-$   &6119                 &6025                 &?5970              &6209                 &6127                 &--\\
$1\,\,{\rm D}\,\,\frac{3}{2}\,\,2^-$   &6121                 &6037                 &                   &6218                 &6140                 &--\\
$1\,\,{\rm D}\,\,\frac{5}{2}\,\,2^-$   &6103                 &5985                 &                   &6189                 &6095                 &--\\
$1\,\,{\rm D}\,\,\frac{5}{2}\,\,3^-$   &6091                 &5993                 &?5970              &6191                 &6103                 &--\\ \hline\hline
\end{tabular}
\end{center}
\caption{The masses of the bottom mesons from two typical potential
models compared to the experimental data.}
\end{table}

The article is arranged as follows:  we study the two-body strong decays of the
bottom mesons $B_1(5721)$,   $B_2(5747)$, $B_{s1}(5830)$, $B_{s2}(5840)$ and $B(5970)$   with the heavy meson
effective theory in Sect.2; in Sect.3, we present the
 numerical results and discussions; and Sect.4 is reserved for our
conclusions.

\section{ The strong  decays with the heavy meson effective theory }
In the heavy quark limit, the heavy-light  mesons
$Q{\bar q}$  can be  classified in doublets according to the total
angular momentum of the light antiquark ${\vec s}_\ell$,
${\vec s}_\ell= {\vec s}_{\bar q}+{\vec L} $, where the ${\vec
s}_{\bar q}$ and ${\vec L}$ are the spin and orbital angular momentum of the light antiquark, respectively \cite{RevWise}.
The doublet $(P,P^*)$ have the spin-parity
$J^P_{s_\ell}=(0^-,1^-)_{\frac{1}{2}}$ for $L=0$ (S-wave);
 the two doublets $(P^*_0,P_1)$ and $(P_1,P^*_2)$ have the spin-parity
$J^P_{s_\ell}=(0^+,1^+)_{\frac{1}{2}}$ and $(1^+,2^+)_{\frac{3}{2}}$
respectively for $L=1$ (P-wave);  the  two doublets $(P^*_1,P_2)$ and
$(P_2,P_3^{ *})$ have the spin-parity
$J^P_{s_\ell}=(1^-,2^-)_{\frac{3}{2}}$ and $(2^-,3^-)_{\frac{5}{2}}$
respectively for $L=2$ (D-wave).
In the heavy meson effective theory,  those  doublets with the same radial quantum numbers can be
described by the effective super-fields $H_a$, $S_a$,  $T_a$, $X_a$ and $Y_a$,  respectively \cite{Falk1992},
\begin{eqnarray}
H_a & =& \frac{1+{\rlap{v}/}}{2}\left\{P_{a\mu}^*\gamma^\mu-P_a\gamma_5\right\} \, ,   \nonumber  \\
S_a &=& \frac{1+{\rlap{v}/}}{2} \left\{P_{1a}^{ \mu}\gamma_\mu\gamma_5-P_{0a}^*\right\}  \, , \nonumber \\
T_a^\mu &=&\frac{1+{\rlap{v}/}}{2} \left\{ P^{*\mu\nu}_{2a}\gamma_\nu-P_{1a\nu} \sqrt{3 \over 2} \gamma_5 \left[ g^{\mu \nu}-{\gamma^\nu
(\gamma^\mu-v^\mu) \over 3} \right]\right\}\, ,   \nonumber\\
X_a^\mu &=&\frac{1+{\rlap{v}/}}{2} \Bigg\{ P^{\mu\nu}_{2a}
\gamma_5\gamma_\nu -P^{*}_{1a\nu} \sqrt{3 \over 2} \left[ g^{\mu \nu}-{\gamma^\nu (\gamma^\mu+v^\mu) \over 3}  \right]\Bigg\} \, , \nonumber  \\
Y_a^{ \mu\nu} &=&\frac{1+{\rlap{v}/}}{2} \left\{P^{*\mu\nu\sigma}_{3a} \gamma_\sigma -P^{\alpha\beta}_{2a}\sqrt{5 \over 3} \gamma_5 \left[ g^\mu_\alpha g^\nu_\beta -{g^\nu_\beta\gamma_\alpha  (\gamma^\mu-v^\mu) \over 5} - {g^\mu_\alpha\gamma_\beta  (\gamma^\nu-v^\nu) \over 5}  \right]\right\}\, ,
\end{eqnarray}
where the  heavy meson fields  $P^{(*)}$ contain a factor $\sqrt{M_{P^{(*)}}}$ and
have dimension of mass $\frac{3}{2}$.
The super-fields $H_a$ contain the $\rm{S}$-wave mesons $(P,P^*)$; $S_a$, $T_a$ contain  the $\rm{P}$-wave mesons $(P^*_0,P_1)$, $(P_1,P^*_2)$ respectively; $X_a$, $Y_a$ contain  the $\rm{D}$-wave mesons $(P^*_1,P_2)$,
$(P_2,P_3^{ *})$ respectively.

 The light pseudoscalar mesons are described by the fields
 $\displaystyle \xi=e^{i {\cal M} \over
f_\pi}$, where
\begin{equation}
{\cal M}= \left(\begin{array}{ccc}
\sqrt{\frac{1}{2}}\pi^0+\sqrt{\frac{1}{6}}\eta & \pi^+ & K^+\nonumber\\
\pi^- & -\sqrt{\frac{1}{2}}\pi^0+\sqrt{\frac{1}{6}}\eta & K^0\\
K^- & {\bar K}^0 &-\sqrt{\frac{2}{3}}\eta
\end{array}\right) \, ,
\end{equation}
and $f_\pi=130\,\rm{MeV}$.

At the leading order approximation, the heavy meson chiral Lagrangians ${\cal L}_0$, ${\cal
L}_{HH}$, ${\cal L}_{SH}$, ${\cal L}_{TH}$, ${\cal L}_{XH}$, ${\cal L}_{XS}$, ${\cal L}_{XT}$,  ${\cal L}_{YH}$, ${\cal L}_{YS}$, ${\cal L}_{YT}$  for
the two-body strong decays to the light pseudoscalar mesons  can be  written as:
\begin{eqnarray}
{\cal L}_0&=&i{\rm Tr} \left\{{\bar H}_a {v\cdot\cal D}_{ab} H_b     \right\} +i{\rm Tr} \left\{{\bar S}_a {v\cdot\cal D}_{ab} S_b\right\}
+i{\rm Tr} \left\{{\bar T}^\mu_a {v\cdot\cal D}_{ab} T_{\mu b}   \right\} +i{\rm Tr} \left\{{\bar X}^\mu_a {v\cdot\cal D}_{ab} X_{\mu b}   \right\}
\nonumber \\
 &&+i{\rm Tr} \left\{{\bar Y}^{\mu\nu}_a {v\cdot\cal D}_{ab} Y_{\mu\nu b}   \right\}-\delta m_S{\rm Tr} \left\{{\bar S}_a  S_a\right\}-\delta m_T{\rm Tr} \left\{{\bar T}^\mu_a  T_{\mu a}   \right\} -\delta m_X {\rm Tr} \left\{{\bar X}^\mu_a X_{\mu a}   \right\}\nonumber\\
 &&-\delta m_Y{\rm Tr} \left\{{\bar Y}^{\mu\nu}_a  Y_{\mu\nu a}   \right\}\, ,\nonumber \\
{\cal L}_{HH} &=&   g_{HH} {\rm Tr} \left\{{\bar H}_a H_b \gamma_\mu\gamma_5 {\cal A}_{ba}^\mu \right\} \, ,\nonumber \\
{\cal L}_{SH} &=&   g_{SH} {\rm Tr} \left\{{\bar H}_a S_b \gamma_\mu \gamma_5 {\cal A}_{ba}^\mu \right\}\, + \, h.c. \, , \nonumber \\
{\cal L}_{TH} &=&  {g_{TH} \over \Lambda}{\rm Tr}\left\{{\bar H}_aT^\mu_b (i {\cal D}_\mu {\not\! {\cal A}  }+i{\not\! {\cal D}  } { \cal A}_\mu)_{ba} \gamma_5\right\} + h.c. \, , \nonumber \\
{\cal L}_{XH} &=& {g_{XH} \over \Lambda}{\rm Tr}\left\{{\bar H}_a X^\mu_b(i {\cal D}_\mu {\not\! {\cal A}  }+i{\not\! {\cal D}  } { \cal A}_\mu)_{ba} \gamma_5\right\} + h.c. \, ,\nonumber   \\
{\cal L}_{XS} &=& {g_{XS} \over \Lambda}{\rm Tr}\left\{{\bar S}_a X^\mu_b(i {\cal D}_\mu {\not\! {\cal A}  }+i{\not\! {\cal D}  } { \cal A}_\mu)_{ba} \gamma_5\right\} + h.c. \, ,\nonumber   \\
{\cal L}_{XT} &=&  {1 \over {\Lambda^2}}{\rm Tr}\left\{ {\bar T}^\mu_a X^{ \nu}_b \left[k_1^T \{{\cal D}_\mu, {\cal D}_\nu\} {\cal A}_\lambda + k_2^T \left({\cal D}_\mu {\cal D}_\lambda { \cal A}_\nu + {\cal D}_\nu {\cal D}_\lambda { \cal A}_\mu \right)\right]_{ba}  \gamma^\lambda \gamma_5\right\} + h.c. \, ,\nonumber   \\
{\cal L}_{YH} &=&  {1 \over {\Lambda^2}}{\rm Tr}\left\{ {\bar H}_a Y^{\mu \nu}_b \left[k_1^H \{{\cal D}_\mu, {\cal D}_\nu\} {\cal A}_\lambda + k_2^H \left({\cal D}_\mu {\cal D}_\lambda { \cal A}_\nu + {\cal D}_\nu {\cal D}_\lambda { \cal A}_\mu \right)\right]_{ba}  \gamma^\lambda \gamma_5\right\} + h.c. \, ,\nonumber   \\
{\cal L}_{YS} &=&  {1 \over {\Lambda^2}}{\rm Tr}\left\{ {\bar S}_a Y^{\mu \nu}_b \left[k_1^S \{{\cal D}_\mu, {\cal D}_\nu\} {\cal A}_\lambda + k_2^S \left({\cal D}_\mu {\cal D}_\lambda { \cal A}_\nu + {\cal D}_\nu {\cal D}_\lambda { \cal A}_\mu \right)\right]_{ba}  \gamma^\lambda \gamma_5\right\} + h.c. \, ,\nonumber   \\
{\cal L}_{YT} &=& {g_{YT} \over \Lambda}{\rm Tr}\left\{{\bar T}_{a\mu} X^{\mu\nu}_b(i {\cal D}_\nu {\not\! {\cal A}  }+i{\not\! {\cal D}  } { \cal A}_\nu)_{ba} \gamma_5\right\} + h.c. \, ,
  \end{eqnarray}
where
\begin{eqnarray}
{\cal D}_{\mu}&=&\partial_\mu+{\cal V}_{\mu} \, , \nonumber \\
 {\cal V}_{\mu }&=&\frac{1}{2}\left(\xi^\dagger\partial_\mu \xi+\xi\partial_\mu \xi^\dagger\right)\, , \nonumber \\
 {\cal A}_{\mu }&=&\frac{1}{2}\left(\xi^\dagger\partial_\mu \xi-\xi\partial_\mu  \xi^\dagger\right)\,  , \nonumber\\
 \{ {\cal D}_\mu, {\cal D}_\nu \}&=&{\cal D}_\mu {\cal D}_\nu+{\cal D}_\nu {\cal D}_\mu  \, ,
\end{eqnarray}
  $\delta m_S=m_S-m_H$, $\delta m_T=m_T-m_H$,  $\delta m_X=m_X-m_H$, $\delta m_Y=m_Y-m_H$, $\Lambda$ is the chiral symmetry-breaking scale and chosen as
$\Lambda = 1 \, \rm{GeV} $  \cite{Colangelo0511}, the hadronic coupling
constants  $g_{HH}$, $g_{SH}$, $g_{TH}$, $g_{XH}$, $g_{XS}$, $g_{XT}=k^T_1+k^T_2$, $g_{YH}=k^H_1+k^H_2$, $g_{YS}=k^S_1+k^S_2$  and $g_{YT}$ depend on the radial quantum numbers of the heavy mesons, and  can be
fitted  to the experimental data, if they are available.
The heavy meson chiral Lagrangians  ${\cal
L}_{HH}$, ${\cal L}_{SH}$, ${\cal L}_{TH}$, ${\cal L}_{XH}$ and ${\cal L}_{YH}$  are taken from Ref.\cite{HL-1}, the  ${\cal L}_{XS}$, ${\cal L}_{XT}$, ${\cal L}_{YS}$ and ${\cal L}_{YT}$
are constructed accordingly  in this article.  The flavor and spin violation corrections of order
$\mathcal {O}(1/m_Q)$ are neglected,  we expect that the corrections are not  larger
than (or as large as) the leading order contributions.

From the heavy meson chiral Lagrangians  ${\cal L}_{HH}$, ${\cal L}_{SH}$, ${\cal L}_{TH}$, ${\cal L}_{XH}$, ${\cal L}_{XS}$, ${\cal L}_{XT}$,  ${\cal L}_{YH}$, ${\cal L}_{YS}$, ${\cal L}_{YT}$, we can obtain the  widths
$\Gamma$ of the two-body strong decays to the light pseudoscalar mesons,
\begin{eqnarray}
\Gamma&=&\frac{1}{2J+1}\sum\frac{p_{f}}{8\pi M^2_i } |{\cal A}|^2\, , \nonumber\\
p_f&=&\frac{\sqrt{(M_i^2-(M_f+m_\mathcal{P})^2)(M_i^2-(M_f-m_\mathcal{P})^2)}}{2M_i}\, ,
\end{eqnarray}
where the ${\cal A}$ denotes the scattering amplitudes,  the $i$ and $f$ denote the initial and final state heavy mesons, respectively, the  $J$ is the total angular momentum  of the initial heavy meson, the $\sum$ denotes the summation of all the  polarization vectors, and the $\mathcal{P}$ denotes the light pseudoscalar mesons.

Now we write down the explicit expressions of the decay widths $\Gamma$ in different channels. There are 53 expressions, 14 expressions are  originally  obtained in previous works by other authors \cite{Colangelo1207,Colangelo0607,PRT1997}, while 39 expressions are obtained in this article.

$\bullet$  $(0^-,1^-)_{\frac{1}{2}}\to (0^-,1^-)_{\frac{1}{2}}+ \mathcal{P}$,
\begin{eqnarray}
\Gamma(1^- \to 1^-) &=&C_{\mathcal{P}} \frac{g_{HH}^2M_f p_f^3}{3\pi f_{\pi}^2M_i} \, , \\
\Gamma(1^- \to 0^-) &=&C_{\mathcal{P}} \frac{g_{HH}^2M_f p_f^3}{6\pi f_{\pi}^2M_i}\, , \\
\Gamma(0^- \to 1^-) &=&C_{\mathcal{P}} \frac{g_{HH}^2M_f p_f^3}{2\pi f_{\pi}^2M_i}  \, ,
\end{eqnarray}
which take place through relative P-wave,  the experimental candidates are $B(5970)^0({\rm 2 S}) \to B^+\pi^-$, $B(5970)^+({\rm 2 S}) \to B^0 \pi^+$ \cite{CDF-2013}.
The three expressions   differ from that obtained in early works by some factors \cite{PRT1997}, while they are consistent with that obtained in Ref.\cite{Colangelo1207}, as the  same conventions  are taken in the present work and in Refs.\cite{Wang1009,Wang1308,Colangelo1207,Colangelo0511,Colangelo0607,Colangelo0710,Colangelo1001}.

$\bullet$  $(0^+,1^+)_{\frac{1}{2}}\to (0^-,1^-)_{\frac{1}{2}}+\mathcal{P}$,
\begin{eqnarray}
\Gamma(1^+ \to 1^-) &=&C_{\mathcal{P}} \frac{g_{SH}^2M_f \left(p_f^2 + m_\mathcal{P}^2 \right)p_f}{2\pi f_{\pi}^2 M_i } \, , \\
\Gamma(0^+ \to 0^-) &=&C_{\mathcal{P}} \frac{g_{SH}^2M_f \left(p_f^2 + m_\mathcal{P}^2 \right)p_f}{2\pi f_{\pi}^2 M_i } \, ,
\end{eqnarray}
which take place through relative S-wave, the $\rm {1P}$ $(0^+,1^+)_{\frac{1}{2}}$ states are expected to be broad, no experimental candidate exists at present time.
The two expressions   differ from that obtained in early works by some factors \cite{PRT1997}, while they are consistent with that in Ref.\cite{Colangelo1207}.

$\bullet$  $(0^-,1^-)_{\frac{1}{2}}\to (0^+,1^+)_{\frac{1}{2}}+\mathcal{P}$,
\begin{eqnarray}
\Gamma(1^- \to 1^+) &=&C_{\mathcal{P}} \frac{g_{SH}^2M_f \left(p_f^2 + m_\mathcal{P}^2 \right)p_f}{2\pi f_{\pi}^2 M_i } \, , \\
\Gamma(0^- \to 0^+) &=&C_{\mathcal{P}} \frac{g_{SH}^2M_f \left(p_f^2 + m_\mathcal{P}^2 \right)p_f}{2\pi f_{\pi}^2 M_i } \, ,
\end{eqnarray}
which take place through relative S-wave,  no experimental candidate exists at present time.

$\bullet$  $(1^+,2^+)_{\frac{3}{2}}\to (0^-,1^-)_{\frac{1}{2}}+ \mathcal{P}$,
\begin{eqnarray}
\Gamma(2^+ \to 1^-) &=&C_{\mathcal{P}} \frac{2g_{TH}^2M_f p_f^5}{5\pi f_{\pi}^2\Lambda^2 M_i} \, , \\
\Gamma(2^+ \to 0^-) &=&C_{\mathcal{P}} \frac{4g_{TH}^2M_f p_f^5}{15\pi f_{\pi}^2\Lambda^2 M_i} \, ,\\
\Gamma(1^+ \to 1^-) &=&C_{\mathcal{P}} \frac{2g_{TH}^2M_f p_f^5}{3\pi f_{\pi}^2 \Lambda^2 M_i } \, ,
\end{eqnarray}
which take place through relative D-wave, the experimental candidates are $B_1(5721)^0({\rm{1P}})\to B^{*+}\pi^-$, $B_2(5747)^0({\rm{1P}}) \to B^{*+}\pi^-,\,B^{+}\pi^-$ \cite{D0-2007,CDF-2008},
$B_{s1}(5830)^0({\rm{1P}}) \to B^{*+}K^-$ \cite{CDF-Bs-2007,D0-Bs-2007,LHCb-2012}, $B_{s2}^*(5840)^0({\rm{1P}}) \to B^{+}K^-$ \cite{CDF-Bs-2007,D0-Bs-2007,LHCb-2012}, $B_{s2}^*(5840)^0({\rm{1P}}) \to B^{*+}K^-$ \cite{LHCb-2012}.
The three expressions   differ from that obtained in early works by some factors \cite{PRT1997}, while they are consistent with that in Ref.\cite{Colangelo1207}.

$\bullet$  $(0^-,1^-)_{\frac{1}{2}} \to  (1^+,2^+)_{\frac{3}{2}} + \mathcal{P}$,
\begin{eqnarray}
\Gamma(1^- \to 2^+) &=&C_{\mathcal{P}} \frac{2g_{TH}^2M_f p_f^5}{3\pi f_{\pi}^2\Lambda^2 M_i} \, , \\
\Gamma(1^- \to 1^+) &=&C_{\mathcal{P}} \frac{2g_{TH}^2M_f p_f^5}{3\pi f_{\pi}^2\Lambda^2 M_i} \, ,\\
\Gamma(0^- \to 2^+) &=&C_{\mathcal{P}} \frac{4g_{TH}^2M_f p_f^5}{3\pi f_{\pi}^2 \Lambda^2 M_i } \, ,
\end{eqnarray}
which take place through relative D-wave, no experimental candidate exists at present time. The phase-spaces  of the decays $B(5970)^0({\rm 2 S}) \to B_1^{+}\pi^-, \, B_2^{*+}\pi^-$, $B(5970)^+({\rm 2 S}) \to B_1^{0} \pi^+,\,B_2^{*0} \pi^+$ are very small.

$\bullet$  $(1^-,2^-)_{\frac{3}{2}}\to (0^-,1^-)_{\frac{1}{2}}+ \mathcal{P}$,
\begin{eqnarray}
\Gamma(2^- \to 1^-) &=&C_{\mathcal{P}} \frac{2g_{XH}^2M_f\left(p_f^2+m_\mathcal{P}^2\right) p_f^3}{3\pi f_{\pi}^2\Lambda^2 M_i } \, , \\
\Gamma(1^- \to 1^-) &=&C_{\mathcal{P}} \frac{2g_{XH}^2M_f\left(p_f^2+m_\mathcal{P}^2\right) p_f^3}{9\pi f_{\pi}^2\Lambda^2 M_i } \, , \\
\Gamma(1^- \to 0^-) &=&C_{\mathcal{P}} \frac{4g_{XH}^2M_f\left(p_f^2+m_\mathcal{P}^2\right) p_f^3}{9\pi f_{\pi}^2\Lambda^2 M_i } \, ,
\end{eqnarray}
 which take place through relative P-wave, the experimental candidates are $B(5970)^0({\rm 1D}) \to B^+\pi^-$, $B(5970)^+({\rm 1D}) \to B^0\pi^+$ \cite{CDF-2013}.
The three expressions    are consistent with that originally obtained in Ref.\cite{Colangelo1207}.

$\bullet$  $(0^-,1^-)_{\frac{1}{2}} \to (1^-,2^-)_{\frac{3}{2}} + \mathcal{P}$,
\begin{eqnarray}
\Gamma(1^- \to 2^-) &=&C_{\mathcal{P}} \frac{10g_{XH}^2M_f\left(p_f^2+m_\mathcal{P}^2\right) p_f^3}{9\pi f_{\pi}^2\Lambda^2 M_i } \, , \\
\Gamma(1^- \to 1^-) &=&C_{\mathcal{P}} \frac{2g_{XH}^2M_f\left(p_f^2+m_\mathcal{P}^2\right) p_f^3}{9\pi f_{\pi}^2\Lambda^2 M_i } \, , \\
\Gamma(0^- \to 1^-) &=&C_{\mathcal{P}} \frac{4g_{XH}^2M_f\left(p_f^2+m_\mathcal{P}^2\right) p_f^3}{3\pi f_{\pi}^2\Lambda^2 M_i } \, ,
\end{eqnarray}
which take place through relative P-wave, no experimental candidate exists at present time.

$\bullet$  $(1^-,2^-)_{\frac{3}{2}}\to (0^+,1^+)_{\frac{1}{2}}+ \mathcal{P}$,
\begin{eqnarray}
\Gamma(2^- \to 1^+) &=&C_{\mathcal{P}} \frac{2g_{XS}^2M_f p_f^5}{5\pi f_{\pi}^2\Lambda^2 M_i } \, , \\
\Gamma(2^- \to 0^+) &=&C_{\mathcal{P}} \frac{4g_{XS}^2M_f p_f^5}{15\pi f_{\pi}^2\Lambda^2 M_i } \, , \\
\Gamma(1^- \to 1^+) &=&C_{\mathcal{P}} \frac{2g_{XS}^2M_f p_f^5}{3\pi f_{\pi}^2\Lambda^2 M_i } \, ,
\end{eqnarray}
which take place through relative D-wave, no experimental candidate exists at present time.

$\bullet$  $(0^+,1^+)_{\frac{1}{2}} \to (1^-,2^-)_{\frac{3}{2}} + \mathcal{P}$,
\begin{eqnarray}
\Gamma(1^+ \to 2^-) &=&C_{\mathcal{P}} \frac{2g_{XS}^2M_f p_f^5}{3\pi f_{\pi}^2\Lambda^2 M_i } \, , \\
\Gamma(1^+ \to 1^-) &=&C_{\mathcal{P}} \frac{2g_{XS}^2M_f p_f^5}{3\pi f_{\pi}^2\Lambda^2 M_i } \, , \\
\Gamma(0^+ \to 2^-) &=&C_{\mathcal{P}} \frac{4g_{XS}^2M_f p_f^5}{3\pi f_{\pi}^2\Lambda^2 M_i } \, ,
\end{eqnarray}
which take place through relative D-wave, no experimental candidate exists at present time.

$\bullet$  $(1^-,2^-)_{\frac{3}{2}}\to (1^+,2^+)_{\frac{3}{2}}+ \mathcal{P}$,
\begin{eqnarray}
\Gamma(2^- \to 2^+) &=&C_{\mathcal{P}} \frac{17g_{XT}^2M_f \left(p_f^2+m_\mathcal{P}^2\right) p_f^5}{45\pi f_{\pi}^2\Lambda^4 M_i } \, , \\
\Gamma(2^- \to 1^+) &=&C_{\mathcal{P}} \frac{g_{XT}^2M_f \left(p_f^2+m_\mathcal{P}^2\right) p_f^5}{15\pi f_{\pi}^2\Lambda^4 M_i } \, , \\
\Gamma(1^- \to 2^+) &=&C_{\mathcal{P}} \frac{g_{XT}^2M_f \left(p_f^2+m_\mathcal{P}^2\right) p_f^5}{9\pi f_{\pi}^2\Lambda^4 M_i } \, , \\
\Gamma(1^- \to 1^+) &=&C_{\mathcal{P}} \frac{g_{XT}^2M_f \left(p_f^2+m_\mathcal{P}^2\right) p_f^5}{3\pi f_{\pi}^2\Lambda^4 M_i } \, ,
\end{eqnarray}
which take place through relative D-wave, no experimental candidate exists at present time. The phase-spaces  of the decays $B(5970)^0({\rm 1 D}) \to B_1^{+}\pi^-, \, B_2^{*+}\pi^-$, $B(5970)^+({\rm 1 D}) \to B_1^{0} \pi^+,\,B_2^{*0} \pi^+$ are very small.

$\bullet$  $(1^+,2^+)_{\frac{3}{2}} \to (1^-,2^-)_{\frac{3}{2}} + \mathcal{P}$,
\begin{eqnarray}
\Gamma(2^+ \to 2^-) &=&C_{\mathcal{P}} \frac{17g_{XT}^2M_f \left(p_f^2+m_\mathcal{P}^2\right) p_f^5}{45\pi f_{\pi}^2\Lambda^4 M_i } \, , \\
\Gamma(2^+ \to 1^-) &=&C_{\mathcal{P}} \frac{g_{XT}^2M_f \left(p_f^2+m_\mathcal{P}^2\right) p_f^5}{15\pi f_{\pi}^2\Lambda^4 M_i } \, , \\
\Gamma(1^+ \to 2^-) &=&C_{\mathcal{P}} \frac{g_{XT}^2M_f \left(p_f^2+m_\mathcal{P}^2\right) p_f^5}{9\pi f_{\pi}^2\Lambda^4 M_i } \, , \\
\Gamma(1^+ \to 1^-) &=&C_{\mathcal{P}} \frac{g_{XT}^2M_f \left(p_f^2+m_\mathcal{P}^2\right) p_f^5}{3\pi f_{\pi}^2\Lambda^4 M_i } \, ,
\end{eqnarray}
which take place through relative D-wave, no experimental candidate exists at present time.

$\bullet$  $(2^-,3^-)_{\frac{5}{2}}\to (0^-,1^-)_{\frac{1}{2}}+ \mathcal{P}$,
\begin{eqnarray}
\Gamma(3^- \to 1^-) &=&C_{\mathcal{P}} \frac{16g_{YH}^2M_f p_f^7}{105\pi f_{\pi}^2\Lambda^4 M_i}\, , \\
\Gamma(3^- \to 0^-) &=&C_{\mathcal{P}} \frac{4g_{YH}^2M_f p_f^7}{35\pi f_{\pi}^2\Lambda^4 M_i} \, , \\
\Gamma(2^- \to 1^-) &=&C_{\mathcal{P}} \frac{4g_{YH}^2M_f p_f^7}{15\pi f_{\pi}^2\Lambda^4 M_i} \, ,
\end{eqnarray}
which take place through relative F-wave, the experimental candidates are $B(5970)^0({\rm 1D}) \to B^+\pi^-$, $B(5970)^+({\rm 1D}) \to B^0\pi^+$ \cite{CDF-2013}.
The three expressions    are consistent with that originally obtained in Refs.\cite{Colangelo1207,Colangelo0607}.

$\bullet$  $(0^-,1^-)_{\frac{1}{2}} \to (2^-,3^-)_{\frac{5}{2}} + \mathcal{P}$,
\begin{eqnarray}
\Gamma(1^- \to 3^-) &=&C_{\mathcal{P}} \frac{16g_{YH}^2M_f p_f^7}{45\pi f_{\pi}^2\Lambda^4 M_i}\, , \\
\Gamma(1^- \to 2^-) &=&C_{\mathcal{P}} \frac{4g_{YH}^2M_f p_f^7}{9\pi f_{\pi}^2\Lambda^4 M_i} \, , \\
\Gamma(0^- \to 3^-) &=&C_{\mathcal{P}} \frac{4g_{YH}^2M_f p_f^7}{5\pi f_{\pi}^2\Lambda^4 M_i} \, ,
\end{eqnarray}
which take place through relative F-wave, no experimental candidate exists at present time.

$\bullet$  $(2^-,3^-)_{\frac{5}{2}}\to (0^+,1^+)_{\frac{1}{2}}+ \mathcal{P}$,
\begin{eqnarray}
\Gamma(3^- \to 1^+) &=&C_{\mathcal{P}} \frac{4g_{YS}^2M_f \left(p_f^2+m_\mathcal{P}^2\right) p_f^5}{15\pi f_{\pi}^2\Lambda^4 M_i}\, , \\
\Gamma(2^- \to 1^+) &=&C_{\mathcal{P}} \frac{8g_{YS}^2M_f \left(p_f^2+m_\mathcal{P}^2\right) p_f^5}{75\pi f_{\pi}^2\Lambda^4 M_i}\, , \\
\Gamma(2^- \to 0^+) &=&C_{\mathcal{P}} \frac{4g_{YS}^2M_f \left(p_f^2+m_\mathcal{P}^2\right) p_f^5}{25\pi f_{\pi}^2\Lambda^4 M_i}\, ,
\end{eqnarray}
which take place through relative D-wave, no experimental candidate exists at present time.

$\bullet$  $(0^+,1^+)_{\frac{1}{2}} \to (2^-,3^-)_{\frac{5}{2}} + \mathcal{P}$,
\begin{eqnarray}
\Gamma(1^+ \to 3^-) &=&C_{\mathcal{P}} \frac{28g_{YS}^2M_f \left(p_f^2+m_\mathcal{P}^2\right) p_f^5}{45\pi f_{\pi}^2\Lambda^4 M_i}\, , \\
\Gamma(1^+ \to 2^-) &=&C_{\mathcal{P}} \frac{8g_{YS}^2M_f \left(p_f^2+m_\mathcal{P}^2\right) p_f^5}{45\pi f_{\pi}^2\Lambda^4 M_i}\, , \\
\Gamma(0^+ \to 2^-) &=&C_{\mathcal{P}} \frac{4g_{YS}^2M_f \left(p_f^2+m_\mathcal{P}^2\right) p_f^5}{5\pi f_{\pi}^2\Lambda^4 M_i}\, ,
\end{eqnarray}
which take place through relative D-wave, no experimental candidate exists at present time.

$\bullet$  $(2^-,3^-)_{\frac{5}{2}}\to (1^+,2^+)_{\frac{3}{2}}+ \mathcal{P}$,
\begin{eqnarray}
\Gamma(3^- \to 2^+) &=&C_{\mathcal{P}} \frac{4g_{YT}^2M_f   p_f^5}{15\pi f_{\pi}^2\Lambda^2 M_i}\, , \\
\Gamma(3^- \to 1^+) &=&C_{\mathcal{P}} \frac{2g_{YT}^2M_f   p_f^5}{45\pi f_{\pi}^2\Lambda^2 M_i}\, , \\
\Gamma(2^- \to 2^+) &=&C_{\mathcal{P}} \frac{7g_{YT}^2M_f   p_f^5}{75\pi f_{\pi}^2\Lambda^2 M_i}\, , \\
\Gamma(2^- \to 1^+) &=&C_{\mathcal{P}} \frac{49g_{YT}^2M_f   p_f^5}{225\pi f_{\pi}^2\Lambda^2 M_i}\, ,
\end{eqnarray}
which take place through relative D-wave, no experimental candidate exists at present time. The phase-spaces  of the decays $B(5970)^0({\rm 1 D}) \to B_1^{+}\pi^-, \, B_2^{*+}\pi^-$, $B(5970)^+({\rm 1 D}) \to B_1^{0} \pi^+,\,B_2^{*0} \pi^+$ are very small.

$\bullet$  $(1^+,2^+)_{\frac{3}{2}} \to (2^-,3^-)_{\frac{5}{2}}+ \mathcal{P}$,
\begin{eqnarray}
\Gamma(2^+ \to 3^-) &=&C_{\mathcal{P}} \frac{28g_{YT}^2M_f   p_f^5}{75\pi f_{\pi}^2\Lambda^2 M_i}\, , \\
\Gamma(2^+ \to 2^-) &=&C_{\mathcal{P}} \frac{7g_{YT}^2M_f   p_f^5}{75\pi f_{\pi}^2\Lambda^2 M_i}\, , \\
\Gamma(1^+ \to 3^-) &=&C_{\mathcal{P}} \frac{14g_{YT}^2M_f   p_f^5}{135\pi f_{\pi}^2\Lambda^2 M_i}\, , \\
\Gamma(1^+ \to 2^-) &=&C_{\mathcal{P}} \frac{49g_{YT}^2M_f   p_f^5}{135\pi f_{\pi}^2\Lambda^2 M_i}\, ,
\end{eqnarray}
which take place through relative D-wave, no experimental candidate exists at present time.

 The coefficients $C_{\pi^\pm}=C_{K^\pm}=C_{K^0}=C_{\bar{K}^0}=1$, $C_{\pi^0}=\frac{1}{2}$ and $C_{\eta}=\frac{1}{6}$ or $\frac{2}{3}$; the values $C_{\eta}=\frac{1}{6}$ and $\frac{2}{3}$ correspond to the initial states $b\bar{u}$ (or $b\bar{d}$) and $b\bar{s}$ states, respectively. There are minor errors in Ref.\cite{Wang1308}, the numerical values of the decay widths concerning the final state $\eta$ in Table 4-7 should be divided by 4, as the coefficient $C_{\eta}=\frac{2}{3}$ in stead of $C_{\eta}=\frac{1}{6}$ is taken in Ref.\cite{Wang1308}.

\section{Numerical Results and Discussions}
The input parameters are taken  as
$M_{\pi^+}=139.57\,\rm{MeV}$, $M_{\pi^0}=134.9766\,\rm{MeV}$,
$M_{K^+}=493.677\,\rm{MeV}$, $M_{K^0}=497.614\,\rm{MeV}$, $M_{\eta}=547.853\,\rm{MeV}$,
$M_{B^+}=5.27925\,\rm{GeV}$,  $M_{B^0}=5.27955\,\rm{GeV}$,  $M_{B^*}=5.3252\,\rm{GeV}$, $M_{B_1(5721)}=5.7235\,\rm{GeV}$,
$M_{B_2(5747)}=5.743\,\rm{GeV}$, $M_{B_s}=5.3667\,\rm{GeV}$, $M_{B_s^*}=5.4158\,\rm{GeV}$, $M_{B_{s1}(5830)}=5.8287\,\rm{GeV}$,
      $M_{B_{s2}(5840)}=5.83996\,\rm{GeV}$ \cite{PDG}, and
 $M_{B(5970)}=5978\,\rm{MeV}$ \cite{CDF-2013}.

 The numerical values of the decay widths  of  the bottom mesons $B_1(5721)$, $B_{2}(2650)$, $B_{s1}(5830)$, $B_{s2}(5840)$,
 $B(5970)$  are presented in Table 2, where we retain the strong coupling constants $g_{HH}$, $g_{TH}$, $g_{XH}$, $g_{YH}$, $g_{XT}$ and $g_{YT}$.

 The values  $\Gamma_{B_2^*}= \left(22.7^{+3.8}_{-3.2} {}^{+3.2}_{-10.2}\right)\,\rm{MeV}$ and $\Gamma_{B_{s2}^*}=\left(1.56\pm0.13\pm0.47\right)\,\rm{MeV}$ listed in the Review of Particle Physics are taken from the experimental data of   the CDF   collaboration  \cite{CDF-2008} and  LHCb collaboration \cite{LHCb-2012}, respectively.
  Recently, the CDF collaboration measured all the  widths of the $B_{1}^0$, $B_2^{*0}$, $B_1^+$, $B_2^{*+}$, $B_{s1}$ and $B_{s2}^*$ for the first time \cite{CDF-2013}.
  We can saturate the widths of the $B_{1}^0$, $B_2^{*0}$, $B_{s1}$, $B_{s2}^*$ with the two-body strong decays  and confront  the decay widths in Table 2 with the experimental data,
  \begin{eqnarray}
\Gamma_{B_1(5721)^0}&=&0.10683\,g_{TH}^2\,{\rm{GeV}}=(20 \pm 2 \pm 5)\,\rm{MeV}\, \cite{CDF-2013}\, , \\
\Gamma_{B_2(5747)^0}&=&0.17870\,g_{TH}^2\,{\rm{GeV}} =(26 \pm 3 \pm 3)\,{\rm{MeV}}\, \cite{CDF-2013} \, ,\\
\Gamma_{B_{s1}(5830)}&=&0.00011\,g_{TH}^2\,{\rm{GeV}}=(0.7 \pm 0.3 \pm 0.3) \,{\rm{MeV}}\, \cite{CDF-2013} \, ,\\
\Gamma_{B_{s2}(5840)}&=&0.00929\,g_{TH}^2\,{\rm{GeV}}= (2.0 \pm 0.4 \pm 0.2) \,{\rm{MeV}}\, \cite{CDF-2013} \, ,
\end{eqnarray}
to obtain the hadronic coupling constant $g_{TH}$,
\begin{eqnarray}
g_{TH}&=&0.433\pm0.058\,\,\, \,\,\,{\rm from}\,\,\,\,\,\,\Gamma_{B_1(5721)^0} \, ,\\
g_{TH}&=&0.381\pm0.031\,\,\, \,\,\,{\rm from}\,\,\,\,\,\,\Gamma_{B_2(5747)^0}\, ,\\
g_{TH}&=&0.464\pm0.052\,\,\, \,\,\,{\rm from}\,\,\,\,\,\,\Gamma_{B_{s2}(5840)}\, ,
\end{eqnarray}
where we  neglect the large value $g_{TH}=2.52$ from the small decay width $\Gamma_{B_{s1}(5830)}$, the strong decays of the $B_{s1}(5830)$ are greatly depressed in the phase-space.
The average value is
\begin{eqnarray}
g_{TH}&=&0.43\pm0.05\, ,
\end{eqnarray}
which is consistent with the value $h^{\prime}=0.43\pm0.01$ extracted from the decays of the charmed mesons \cite{Colangelo1207}. The heavy quark symmetry works well,
a universal  hadronic coupling constant $g_{TH}$ (or $h^{\prime}$) exists. We extract the hadronic coupling constant $g_{TH}$ in the bottom sector for the first time.

The $B(5970)$ have three possible assignments: the $2{\rm S}\,1^-$, $1{\rm D}\,1^-$ and $1{\rm D}\,3^-$ states, the corresponding two-body strong decays are quite different,
the numerical values of the decay widths are shown explicitly in Table 2.
From Table 2, we can obtain  the ratios  among the partial decay widths so as to identify the $B(5970)$ by confronting them with the experimental data in the future, as
the  hadronic coupling constants are canceled out.
Again, we can saturate the widths with the two-body strong decays  to the S-wave ground mesons, as the decays to the P-wave mesons are greatly depressed in the phase-space, and confront  the widths with  the experimental data from  the CDF   collaboration \cite{CDF-2013},
\begin{eqnarray}
\Gamma_{B(5970)^0}(2{\rm S}\,1^-)&=&3.21730\,g_{HH}^2\,{\rm{GeV}}=(70 \pm 18 \pm 31)\,\rm{MeV}\, \cite{CDF-2013}\, , \\
\Gamma_{B(5970)^0}(1{\rm D}\,1^-)&=&1.93262\,g_{XH}^2\,{\rm{GeV}}=(70 \pm 18 \pm 31)\,\rm{MeV}\, \cite{CDF-2013}\, , \\
\Gamma_{B(5970)^0}(1{\rm D}\,3^-)&=&0.24541\,g_{YH}^2\,{\rm{GeV}}=(70 \pm 18 \pm 31)\,\rm{MeV}\, \cite{CDF-2013}\, ,
\end{eqnarray}
to obtain the hadronic coupling constants,
\begin{eqnarray}
g_{HH}&=&0.148\pm 0.038\, , \\
g_{XH}&=&0.190\pm 0.049\, , \\
g_{YH}&=&0.534\pm 0.137\, .
\end{eqnarray}
We can take those hadronic coupling constants as basic input parameters and calculate the partial decay widths. The  numerical values of the  partial decay widths are shown explicitly in Table 3, which can be directly confronted with the experimental data  from the LHCb,  CDF, D0 and KEK-B  collaborations in the future to identify the $B(5970)$.

We can also study the decays to the light vector mesons $V$ besides the pseudoscalar mesons $\mathcal{P}$ with the  replacement ${\cal V}_{\mu} \to {\cal V}_{\mu}+V_\mu$,  and introduce additional phenomenological Lagrangians \cite{PRT1997}, therefore additional unknown
coupling constants. The decays to the vector mesons $V$ are depressed in the phase-space compared to the light pseudoscalar mesons, we prefer to study those decays  when the experimental data are accumulated. On the other hand, we can also study those strong decays with the chiral quark models \cite{ZhongXH}.

\begin{table}
\begin{center}
\begin{tabular}{|c|c|cc|cc| }\hline\hline
                   & $n\,L\,s_\ell\,J^P$         & Decay channels    & Widths  [GeV]      &Decay channels     & Widths [GeV]   \\ \hline

 $B_1(5721)$       & $1\,P\,\frac{3}{2}\,1^+$    & $B^{*+}\pi^-$     & $0.07068\,g_{TH}^2$    &                  &      \\
                   &                             & $B^{*0}\pi^0$     & $0.03615\,g_{TH}^2$    &                  &       \\ \hline

$B_{2}^*(5747)$    & $1\,P\,\frac{3}{2}\,2^+$    & $B^{*+}\pi^-$     & $0.05495\,g_{TH}^2$    & $B^{+}\pi^-$     & $0.06353\,g_{TH}^2$      \\
                   &                             & $B^{*0}\pi^0$     & $0.02804\,g_{TH}^2$    & $B^{0}\pi^0$     & $0.03218\,g_{TH}^2$      \\ \hline

$B_{s1}(5830)$     & $1\,P\,\frac{3}{2}\,1^+$    & $B^{*+}K^-$       & $0.00009\,g_{TH}^2$    &                  &      \\
                   &                             & $B^{*0}\bar{K}^0$ & $0.00002\,g_{TH}^2$    &                  &      \\ \hline

$B_{s2}^*(5840)$   & $1\,P\,\frac{3}{2}\,2^+$    & $B^{*+}K^-$       & $0.00036\,g_{TH}^2$    & $B^{+}K^-$       & $0.00469\,g_{TH}^2$      \\
                   &                             & $B^{*0}\bar{K}^0$ & $0.00022\,g_{TH}^2$    & $B^{0}\bar{K}^0$ & $0.00402\,g_{TH}^2$        \\ \hline

 $B(5970)$         & $2\,S\,\frac{1}{2}\,1^-$    & $B^{*+}\pi^-$     & $1.22526\,g_{HH}^2$    & $B^{+}\pi^-$     & $0.74252\,g_{HH}^2$   \\
                   &                             & $B_s^{*}K^0$      & $0.08790\,g_{HH}^2$    & $B_s K^0$        & $0.10746\,g_{HH}^2$   \\
                   &                             & $B^{*0}\pi^0$     & $0.61549\,g_{HH}^2$    & $B^{0}\pi^0$     & $0.37232\,g_{HH}^2$   \\
                   &                             & $B^{*0}\eta$      & $0.03510\,g_{HH}^2$    & $B^{0}\eta$      & $0.03125\,g_{HH}^2$   \\
                   &                             & $B_2^{*+}\pi^-$   & $0.00264\,g_{TH}^2$    & $B_1^{+}\pi^-$   & $0.00471\,g_{TH}^2$     \\
                   &                             & $B_2^{*0}\pi^0$   & $0.00144\,g_{TH}^2$    & $B_{1}^{0}\pi^0$ & $0.00252\,g_{TH}^2$      \\ \hline

 $B(5970)$         & $1\,D\,\frac{3}{2}\,1^-$    & $B^{*+}\pi^-$     & $0.31277\,g_{XH}^2$    & $B^{+}\pi^-$     & $0.86132\,g_{XH}^2$   \\
                   &                             & $B_s^{*}K^0$      & $0.01815\,g_{XH}^2$    & $B_s K^0$        & $0.10342\,g_{XH}^2$   \\
                   &                             & $B^{*0}\pi^0$     & $0.15706\,g_{XH}^2$    & $B^{0}\pi^0$     & $0.43140\,g_{XH}^2$   \\
                   &                             & $B^{*0}\eta$      & $0.00965\,g_{XH}^2$    & $B^{0}\eta$      & $0.03885\,g_{XH}^2$   \\
                   &                             & $B_2^{*+}\pi^-$   & $0.00002\,g_{XT}^2$    & $B_1^{+}\pi^-$   & $0.00015\,g_{XT}^2$        \\
                   &                             & $B_2^{*0}\pi^0$   & $0.00001\,g_{XT}^2$    & $B_{1}^{0}\pi^0$ & $0.00008\,g_{XT}^2$      \\ \hline

 $B(5970)$         & $1\,D\,\frac{5}{2}\,3^-$    & $B^{*+}\pi^-$     & $0.07398\,g_{YH}^2$    & $B^{+}\pi^-$     & $0.08791\,g_{YH}^2$   \\
                   &                             & $B_s^{*}K^0$      & $0.00015\,g_{YH}^2$    & $B_s K^0$        & $0.00095\,g_{YH}^2$   \\
                   &                             & $B^{*0}\pi^0$     & $0.03739\,g_{YH}^2$    & $B^{0}\pi^0$     & $0.04424\,g_{YH}^2$   \\
                   &                             & $B^{*0}\eta$      & $0.00020\,g_{YH}^2$    & $B^{0}\eta$      & $0.00059\,g_{YH}^2$   \\
                   &                             & $B_2^{*+}\pi^-$   & $0.00106\,g_{YT}^2$    & $B_1^{+}\pi^-$   & $0.00031\,g_{YT}^2$        \\
                   &                             & $B_2^{*0}\pi^0$   & $0.00058\,g_{YT}^2$    & $B_{1}^{0}\pi^0$ & $0.00017\,g_{YT}^2$      \\ \hline\hline

\end{tabular}
\end{center}
\caption{ The strong decay widths of the bottom mesons $B_1(5721)$,   $B_2(5747)$, $B_{s1}(5830)$, $B_{s2}(5840)$ and $B(5970)$.    }
\end{table}

\begin{table}
\begin{center}
\begin{tabular}{|c|c|cc|cc| }\hline\hline
                   & $n\,L\,s_\ell\,J^P$         & Decay channels    & Widths  [MeV]          &Decay channels     & Widths [MeV]   \\ \hline

 $B(5970)$         & $2\,S\,\frac{1}{2}\,1^-$    & $B^{*+}\pi^-$     & $26.8\pm13.8$          & $B^{+}\pi^-$     & $16.3\pm8.4$   \\
                   &                             & $B_s^{*}K^0$      & $1.9\pm1.0$            & $B_s K^0$        & $2.4\pm1.2$   \\
                   &                             & $B^{*0}\pi^0$     & $13.5\pm6.9$           & $B^{0}\pi^0$     & $8.2\pm4.2$   \\
                   &                             & $B^{*0}\eta$      & $0.8\pm0.4$            & $B^{0}\eta$      & $0.7\pm0.4$   \\  \hline

 $B(5970)$         & $1\,D\,\frac{3}{2}\,1^-$    & $B^{*+}\pi^-$     & $11.3\pm5.8$           & $B^{+}\pi^-$     & $31.1\pm16.0$   \\
                   &                             & $B_s^{*}K^0$      & $0.7\pm0.3$            & $B_s K^0$        & $3.7\pm1.9$   \\
                   &                             & $B^{*0}\pi^0$     & $5.7\pm2.9$            & $B^{0}\pi^0$     & $15.6\pm8.0$   \\
                   &                             & $B^{*0}\eta$      & $0.3\pm0.2$            & $B^{0}\eta$      & $1.4\pm0.7$   \\  \hline

 $B(5970)$         & $1\,D\,\frac{5}{2}\,3^-$    & $B^{*+}\pi^-$     & $21.1\pm10.8$          & $B^{+}\pi^-$     & $25.1\pm12.9$   \\
                   &                             & $B_s^{*}K^0$      & $<0.1$                 & $B_s K^0$        & $0.3\pm0.1$   \\
                   &                             & $B^{*0}\pi^0$     & $10.7\pm5.5$           & $B^{0}\pi^0$     & $12.6\pm6.5$   \\
                   &                             & $B^{*0}\eta$      & $<0.1$                 & $B^{0}\eta$      & $0.2\pm0.1$   \\  \hline

\end{tabular}
\end{center}
\caption{ The strong decay widths of the bottom mesons  $B(5970)$ with three possible assignments.    }
\end{table}

The heavy-light mesons listed in the  Review of Particle Physics can be classified into the spin doublets in the heavy quark limit, such as \\
the $\rm{1S}$ $(0^-,1^-)_{\frac{1}{2}}$ doublets $(B,B^*)$, $(D,D^*)$, $(B_s,B_s^*)$,  $(D_s,D_s^*)$; \\
the $\rm{1P}$ $(0^+,1^+)_{\frac{1}{2}}$ doublets $(D_0^*(2400),D_1(2430))$, $(D_{s0}^*(2317),D_{s1}(2460))$;  \\
the $\rm{1P}$ $(1^+,2^+)_{\frac{3}{2}}$ doublets $(B_1(5721),B^*_2(5747))$,
 $(D_1(2420),D^*_2(2460))$, $(B_{s1}(5830),B^*_{s2}(5840))$, $(D_{s1}(2536),D^*_{s2}(2573))$ \cite{PDG}; \\
  the $\rm{1D}$ $(1^-,2^-)_{\frac{3}{2}}$ doublets  $(D_J^*(2760/2760),D_J(2740/2750))$ \cite{Wang1308}, $(B(5970),\cdots\cdots)$; \\
the $\rm{1D}$ $(2^-,3^-)_{\frac{5}{2}}$ doublets  $(D_J(2740/2750),D_J^*(2760/2760))$ \cite{Wang1308}, $(\cdots\cdots,D_{sJ}(2860))$ \cite{Colangelo0607}, \\ $(\cdots\cdots,B(5970))$; \\
 the $\rm{2S}$ $(0^-,1^-)_{\frac{1}{2}}$ doublets $(D_J(2580/2550),D_J^*(2650/2600))$ \cite{Wang1009,Wang1308}, $(\cdots\cdots,D_{s1}^*(2700))$ \cite{Colangelo0710}, \\ $(\cdots\cdots,B(5970))$; \\
 the $\rm{2P}$ $(0^+,1^+)_{\frac{1}{2}}$ doublet $(\cdots\cdots,D_{sJ}(3040))$ \cite{Colangelo1001};\\
  the $\rm{2P}$ $(1^+,2^+)_{\frac{3}{2}}$ doublets $(D_{sJ}(3040),\cdots\cdots)$ \cite{Colangelo1001}.\\
  The spin doublets are far from complete, but we expect a complete spectrum of the heavy mesons will be available  from the experimental data of the LHCb, CDF, D0 and Belle-II collaborations in the futures. In this article, we obtain a large number of simple expressions for the Okubo-Zweig-Iizuka allowed  two-body strong decays, and we can use those simple expressions to calculate
   the partial decay widths of the  pseudoscalar meson transitions among the spin doublets in a few minutes.

The mass is a fundamental parameter in describing a hadron (or a meson), however, a hadron (or a meson) cannot be identified  unambiguously by the mass alone,
furthermore, the heavy-light meson masses from different theoretical approaches differ from each other in one way or another, and vary in a rather large range \cite{HQEF-mass,LQCD-mass,Potential-mass,EFG,PE,HQS-mass,HMCh-mass,QCD-string-mass}, we have to resort to  the productions  and decays    to identify the special meson. We can calculate the partial decay widths in different channels using the simple expressions obtained in this article, then calculate  the ratios among diffident channels and confront them with  the experimental data to identify the heavy mesons in the futures.  On the other hand, we can obtain the hadronic coupling constants $g_{HH}$, $g_{SH}$, $g_{TH}$, $\cdots$  from both the charm sector and bottom sector, and examine the heavy quark symmetry. Once the hadronic  coupling constants $g_{HH}$, $g_{SH}$, $g_{TH}$, $\cdots$ are determined unambiguously, we can take them as basic input parameters and perform phenomenological analysis. For example, if the $B(5970)$ is the $2{\rm S}\,1^-$ $b\bar{q}$ state, then the hadronic  coupling constant  $ g_{HH}=0.148\pm 0.038$, we can take it as the basic input parameter to calculate the strong decay widths (also the ratios among the strong decay widths) of the partners $2{\rm S}\,1^-$ $b\bar{s}$ state and $2{\rm S}\,0^-$ $b\bar{q}$, $b\bar{s}$ states, which are expected to be observed  by the LHCb, CDF, D0 and Belle-II collaborations in the futures.

\section{Conclusion}
In this article, we study the two-body strong decays of the bottom mesons with the heavy meson  effective theory in the leading order approximation,
and obtain all the analytical expressions of the widths among the S-wave, P-wave and D-wave bottom mesons. As an application, we
tentatively assign  the  bottom mesons $B(5970)$ as the $2{\rm S}\,1^-$, $1{\rm D}\,1^-$ and $1{\rm D}\,3^-$ states,
calculate the decay widths of the $B_1(5721)$,   $B_2(5747)$, $B_{s1}(5830)$, $B_{s2}(5840)$ and $B(5970)$, and obtain the
 hadronic coupling constants by comparing them to the experimental  data and make predications of the decay widths,
  which can be confronted with the experimental data  from the LHCb,  CDF, D0 and KEK-B  collaborations  in the future to identify the $B(5970)$.

\section*{Acknowledgment}
This  work is supported by National Natural Science Foundation,
Grant Number 11375063, the Fundamental Research Funds for the
Central Universities,  and Natural Science Foundation of Hebei province, Grant Number A2014502017.

\end{document}